\documentclass[11pt,a4paper]{article}
\usepackage{jheppub}

\makeatletter
\gdef\@fpheader{}
\makeatother

\DeclareMathOperator{\tr}{tr}
\DeclareMathOperator{\Tr}{Tr}
\DeclareMathOperator{\Det}{Det}

\usepackage{slashed}
\usepackage{siunitx}

\usepackage{contour}
\contourlength{1.3pt}

\usepackage[compat=1.1.0]{tikz-feynman}

\usetikzlibrary{decorations.markings,arrows.meta}

\tikzfeynmanset{ with arrow/.style = {
   decoration={
     markings,
     mark=at position .55
          with {\arrow{Stealth[scale=1.4]}}
     },
   postaction=decorate}
}

\title{Effective scalaron--photon interaction in $f(R)$ gravity}

\author{Yuri Shtanov}

\affiliation{Bogolyubov Institute for Theoretical Physics, Metrologichna St.\@ 14-b, Kiev 03143, Ukraine} %

\emailAdd{shtanov@bitp.kyiv.ua}

\abstract{We revisit the effective coupling of the scalaron to gauge fields in $f(R)$ gravity minimally coupled to the Standard Model, focusing on the scalaron decay into two photons. Treating the scalaron as an intrinsic component of the Jordan-frame metric, we analyse quantum effects arising from its coupling to the energy--momentum tensor. In this framework, the trace anomaly contributes to the scalaron--gauge boson interaction.

Using Fujikawa's approach, we obtain the trace-anomaly contribution, associated with the transformation of matter fields between the Jordan and Einstein frames, first in QED and then in the full Standard Model. The resulting effective scalaron--photon interaction agrees with direct perturbative calculations that include the trace anomaly and differs from the result obtained when only the classical expression for the trace of the energy--momentum tensor is taken into account in the scalaron interaction. In the limit where the scalaron mass is much smaller than the masses of particles circulating in the loop, the diagrammatic contribution from the classical expression of the trace of the energy--momentum tensor cancels the anomaly-induced term, leading to a vanishing effective coupling and a strong suppression of the scalaron decay rate into photons.

These results clarify the origin of discrepancies in the literature concerning the effective scalaron coupling to massless gauge fields. They stem from inequivalent prescriptions for incorporating quantum loop effects in $f (R)$ gravity, leading to different scalaron--gauge field interactions with direct implications for scalaron dark-matter phenomenology.}

\keywords{Models for Dark Matter, Specific BSM Phenomenology, Anomalies in Field and String Theories}

\arxivnumber{2606.09510}

\begin{document}
 
\maketitle
\flushbottom

\section{Introduction}

Modified $f(R)$ gravity is of interest for several reasons, most notably because of its historic role in inflationary cosmology \cite{Starobinsky:1980te, Vilenkin:1985md} (see also the reviews \cite{Sotiriou:2008rp, DeFelice:2010aj, Nojiri:2010wj}). In addition, it provides a dark-matter candidate without introducing any new fundamental fields into the theory \cite{Capozziello:2006uv, Nojiri:2008nt, Cembranos:2008gj, Cembranos:2015svp, Corda:2011aa, Katsuragawa:2016yir, Katsuragawa:2017wge, Yadav:2018llv, Parbin:2020bpp, KumarSharma:2022qdf, Shtanov:2021uif, Shtanov:2022xew, Shtanov:2024nmf, Shtanov:2025nue}.

Among the various proposals within this framework, we focus on $f(R)$ theories analytic in the neighbourhood of $R = 0$, for which the curvature-squared term plays a central role. The possibility of explaining dark matter in this setting was first explored in \cite{Cembranos:2008gj, Cembranos:2015svp, Katsuragawa:2016yir} and further developed in our works \cite{Shtanov:2021uif, Shtanov:2022xew, Shtanov:2024nmf, Shtanov:2025nue}. In these theories, dark matter is identified with the scalaron, the conformal degree of freedom of the metric, oscillating about the minimum of its effective potential. Assuming minimal coupling of matter to the metric, the scalaron mass is the only essential free parameter.

In theories of this type, the scalaron does not couple to conformally invariant matter at the classical level and, in particular, has no direct interaction with massless gauge fields such as the electromagnetic field. Nevertheless, an effective coupling is generated at one loop, allowing the scalaron to decay into massless gauge bosons. The form of the effective scalaron--photon interaction, which governs the decay of the scalaron into photons, has been the subject of some controversy in the literature. The aim of the present work is to identify the origin of these discrepancies and clarify their physical interpretation.

To identify the dynamical degrees of freedom of $f(R)$ gravity, one usually performs a conformal transformation to the Einstein frame,
\begin{equation} \label{metric}
g_{\mu\nu} = e^{- \phi / M} \widetilde g_{\mu\nu} \, ,
\end{equation}
where $M$ is the Planck mass (see below). In the Einstein frame, the field $\phi$ becomes a canonically normalised scalaron with a potential determined by the function $f(R)$, minimally coupled to Einstein gravity described by the metric $\widetilde g_{\mu\nu}$. To linear order in $\phi/M$, the scalaron couples to the trace of the energy--momentum tensor. The question then arises how this interaction should be treated at the quantum level.

One possible viewpoint is to treat the scalaron as an additional scalar field in the Einstein frame. Matrix elements are then computed using the corresponding classical interaction vertices and Feynman diagrams. Such a formulation respects the standard equivalence theorems associated with field redefinitions. Within this framework, the trace anomaly is not included as a separate contribution to scalaron--gauge boson processes; rather, it is reproduced with the opposite sign in the limit of infinitely heavy particles circulating in the loop, as in the case of the Higgs field. This perspective was adopted in \cite{Watanabe:2010vy, Takeda:2014qma} and in our recent work \cite{Shtanov:2025nue}, and effectively underlies the result of \cite{Cembranos:2008gj}.

An alternative viewpoint treats the scalaron as an intrinsic part of the Jordan-frame metric. This description is more direct, since matter fields couple minimally to it. In this case, matrix elements are computed from the interaction $-\frac12 h_{\mu \nu} T^{\mu \nu}$, where $h_{\mu \nu}$ denotes the metric perturbation. The trace is taken only at the final stage by setting $h_{\mu \nu} = - (\phi / M)\eta_{\mu \nu}$ according to \eqref{metric}. This procedure automatically incorporates the trace anomaly, as first discussed in a general metric context in \cite{Dolgov:1980kp, Dolgov:1981nw}. In QED, the role of trace anomaly was analysed in \cite{Giannotti:2008cv, Armillis:2009pq}, and is well established for dilatonic interactions with the Standard Model \cite{Coriano:2012nm}. In $R^2$ gravity this approach was taken in \cite{Gorbunov:2012ns} and further developed in \cite{Kamada:2019pmx}; we shall reproduce its main features below.

In this approach, the effective interaction takes the form
\begin{equation} \label{anom}
\phi \left[ \left( T^\mu{}_\mu \right)_\text{anom} + \left( T^\mu{}_\mu \right)_\text{class} \right] \, ,
\end{equation}
where the first term represents the trace anomaly and the second generates the standard Feynman diagrams associated with the classical trace of the energy--momentum tensor (or equivalently the classical interaction Lagrangian).

The two viewpoints described above, which differ in their treatment of the trace anomaly, therefore correspond to distinct prescriptions for incorporating quantum loop effects in $f(R)$ gravity. As a result, they lead to different effective scalaron--gauge field interactions and distinct predictions for processes involving massless gauge bosons.

In this paper, we adopt the Jordan-frame metric viewpoint and re-examine the effective coupling of the scalaron to gauge fields, focusing in particular on the electromagnetic field. The trace-anomaly contribution at one loop can be consistently derived using Fujikawa's method, via the functional Jacobians associated with the transformation of matter fields from the Jordan to the Einstein frame. To illustrate the procedure, we first consider quantum electrodynamics (QED) coupled to $f(R)$ gravity, and then extend the analysis to the full Standard Model (SM). We revisit the scalaron decay rate into two photons, a process relevant for the phenomenology of scalaron dark matter and its potential detection.

The paper is organised as follows. In section~\ref{sec:scalaron}, we introduce the $f(R)$ gravity model and review the role of the scalaron as a dark matter candidate. In section~\ref{sec:QED}, we derive the effective scalaron--photon coupling in QED\@. In section~\ref{sec:SM}, we generalise the analysis to the Standard Model, compute the effective scalaron coupling to gauge fields, and determine the scalaron decay rate into photons. In section~\ref{sec:frames}, we compare the results of different approaches in the literature. Our conclusions are summarised in section~\ref{sec:summary}. Appendix~\ref{app} details the Euclidean space conventions and the gauge-invariant regularisation of traces within Fujikawa’s method.

\section{Scalaron of $f(R)$ gravity}
\label{sec:scalaron}

In this section, we outline the basic features of the theory. We assume that the gravitational Lagrangian admits a power-series expansion in the scalar curvature $R$:
\begin{equation} \label{Sgs}
L_g = - \frac{M^2}{3} f (R) \, , \qquad f (R) = 2 \Lambda + R - \frac{R^2}{6 m^2} + \ldots \, .
\end{equation} 
Here, $M = \sqrt{3 / 16 \pi G} \approx 2.98 \times 10^{18}\, \text{GeV}$ is a conveniently normalised Planck mass, and $\Lambda \approx \left( 3 \times 10^{-33}\,\text{eV} \right)^2$ is the cosmological constant in the natural units $\hbar = c = 1$. We adopt the metric signature $(+, -, -, -)$, standard in particle physics conventions.

The theory involving terms up to $R^2$ in \eqref{Sgs} corresponds to the Starobinsky inflationary model \cite{Starobinsky:1980te, Vilenkin:1985md}, where setting $m \simeq 10^{-5} M$ yields a primordial power spectrum consistent with current observations \cite{Planck:2018jri}. Even in this context, the quadratic curvature term can hardly be interpreted as a quantum correction to the effective gravitational action arising from the integration of matter degrees of freedom, due to its large coefficient.

When the same model is applied to dark matter rather than inflation, the parameter $m$ must instead lie in the meV--MeV range \cite{Cembranos:2008gj, Cembranos:2015svp}. In this case, $m$ should be regarded as a genuine constant in the gravitational action. The required smallness of $m$ (equivalently, the large dimensionless factor $M^2 / 18 m^2 \sim 10^{41}$--$10^{58}$ multiplying the $R^2$ term) may be viewed in analogy with the extreme smallness of the cosmological constant $\Lambda$ in the gravitational action, whose origin is likewise unknown. 

The additional scalar degree of freedom is identified by performing a transformation from the Jordan frame to the Einstein frame. We first write the action with Lagrangian \eqref{Sgs} in the form
\begin{equation}\label{Sg1}
S_g = - \frac{M^2}{3} \int d^4 x \sqrt{-g}\, \bigl[ \Omega R - U (\Omega) \bigr] \, ,
\end{equation}
where $\Omega$ is a new dimensionless scalar field, and the function $U (\Omega)$ is chosen so that variation with respect to $\Omega$ and its substitution into the action returns the original action: 
\begin{equation} 
U' (\Omega) = R \ \ \Rightarrow \ \  \Omega = \Omega (R) \, , \qquad
f (R) = \bigl[ \Omega R - U (\Omega) \bigr]_{\Omega = \Omega (R)} \, . \end{equation}
Thus, $f (R)$ is the Legendre transform of $U (\Omega)$, and vice versa. 

As a next step, one performs a conformal transformation\footnote{In quantum field theory, this transformation is typically referred to as the Weyl transformation. We use the terminology commonly adopted in the context of gravity theory (see, e.g., \cite{Wald:1984}).} in \eqref{Sg1}:
\begin{equation} \label{om}
g_{\mu\nu} = \Omega^{-1}\, \widetilde g_{\mu\nu} \, , \qquad \Omega = e^{\phi / M} \, ,
\end{equation}
where $\phi$ is a new field (the scalaron) parametrising $\Omega$, and tilded quantities refer to the Einstein-frame metric. In this frame, action \eqref{Sg1} reduces to Einstein gravity coupled to a scalar field (the scalaron) $\phi$. The Lagrangian in the Einstein frame is 
\begin{equation}\label{Sg3}
L_g =  - \frac{M^2}{3} \widetilde R + \frac12 \widetilde g^{\mu\nu} \partial_\mu \phi \partial_\nu \phi - V (\phi) \, ,
\end{equation}
where the scalaron potential $V (\phi)$ is calculated from \eqref{Sg1}:
\begin{equation} \label{V}
V (\phi) = - \frac{M^2}{3} e^{- 2 \phi / M} U \bigl( e^{\phi / M} \bigr) \, .
\end{equation}

It is straightforward to verify that the scalaron potential has extrema, $V'(\phi)=0$, corresponding to Jordan-frame values of $R$ that satisfy
\begin{equation}
R f' (R) = 2 f (R) \, .
\end{equation}
The scalaron mass squared, $m_\phi^2 = V'' (\phi)$, at such an extremum is given by
\begin{equation}
m_\phi^2 = \frac13 \left[ \frac{R}{f'(R)} - \frac{1}{f''(R)} \right] = \frac13 \left[ \frac{R^2}{2 f (R)} - \frac{1}{f''(R)} \right] \, .
\end{equation}
If $m_\phi^2 > 0$, the extremum is a local minimum. From \eqref{V}, one sees that the scalaron potential typically varies on the Planck scale $M$, so that near the minimum it can be well approximated by a quadratic form for field values $|\phi| \ll M$.  

For a small cosmological constant, $\Lambda \ll m^2$, the theory has a local minimum at $\phi / M \approx 4 \Lambda / 3 m^2$, corresponding to $R \approx - 4 \Lambda$ in the Jordan frame, with scalaron mass $m_\phi^2 = m^2 + {\cal O} (\Lambda)$. In what follows, we neglect the small cosmological constant in \eqref{Sgs}, which accounts for dark energy but not for dark matter. In this approximation, the minimum is at $\phi = 0$ (corresponding to $R = 0$), and $m_\phi = m$.

The minimal non-trivial model \cite{Starobinsky:1980te, Vilenkin:1985md} is described by
\begin{equation}\label{fstar}
f (R) = R - \frac{R^2}{6 m^2} \, .
\end{equation}
In the Einstein frame, it produces Lagrangian \eqref{Sg3} with
\begin{equation} \label{Vstar}
V (\phi) = \frac12  m^2 M^2 \left( 1 - e^{- \phi / M} \right)^2 \, . 
\end{equation}
This potential has an infinitely extended plateau at $\phi \gg M$. 

\begin{figure}
\begin{center}
\includegraphics[width=.7\textwidth]{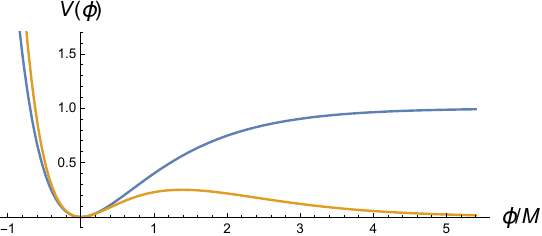}
\caption{Scalaron potentials \eqref{Vstar} (blue) and \eqref{Vrat} (orange) are plotted in units $M^2 m^2/2$. In the region $|\phi|/M \ll 1$, they both are approximated by the quadratic form with mass $m$. \label{fig:pot}}
\end{center}
\end{figure}

Scalaron potentials corresponding to $f(R)$ models with higher powers of $R$ typically exhibit qualitatively different behaviour at large positive $\phi$, including hilltop or tabletop shapes \cite{Shtanov:2022pdx}.  As an example, consider
\begin{equation} 
f (R) = \frac{R}{1 + R / 6 m^2} = R - \frac{R^2}{6 m^2} + \frac{R^3}{36 m^4} - \ldots \, .
\end{equation}
The branch containing the stable critical point $R = 0$ yields the scalaron potential
\begin{equation} \label{Vrat}
V (\phi) = 2 M^2 m^2 e^{- \phi / M}  \left( 1 - e^{- \phi / 2 M} \right)^2 \, .
\end{equation}
This potential has a local maximum at $e^{\phi/M} = 4$ and decreases exponentially as $\phi \to \infty$. Potentials \eqref{Vstar} and \eqref{Vrat} are shown in figure~\ref{fig:pot}.

A common feature of all such potentials in theories containing the quadratic term $- R^2 / 6 m^2$ is that they reduce to a quadratic form with mass $m$ near the origin. The minimal coupling of matter to the metric in the Jordan frame induces a universal scalaron coupling to matter fields, collectively denoted by $\Psi$, with Lagrangian density
\begin{equation}
{\cal L}_\text{m} \left( e^{- \phi / M} \widetilde g_{\mu\nu}, \Psi \right) \, .
\end{equation}
It is the Jordan-frame metric $g_{\mu\nu} = e^{- \phi / M} \widetilde g_{\mu\nu}$ that plays the role of the `observable' metric \cite{Dicke:1961gz, Faraoni:2006fx, Shtanov:2022wpr}. The presence of the scalaron degree of freedom in this metric leads to an additional Yukawa-type gravitational interaction, so that the total gravitational potential per unit gravitating mass is \cite{Stelle:1977ry}
\begin{equation}
\Phi_\text{grav} = - \frac{2 G}{r} \left( 1 + \frac13 e^{- m r} \right) \, .
\end{equation}

The non-observation of Yukawa-type deviations from Newtonian gravity at short distances constrains the range of any gravitational-strength Yukawa interaction to be smaller than 38.6~\SI{}{\micro\metre} at 95\% C.L.\@ \cite{Lee:2020zjt}. This translates into the lower bound
\begin{equation}\label{mlow}
m \geq 5.1~\text{meV} \, .
\end{equation}

If the scalaron constitutes the dark matter component of the universe, its field value in the early Universe must lie very close to the minimum of its potential. Indeed, for the amplitude $\phi_\text{a}$ of the scalaron oscillations one obtains the estimate
\begin{equation}\label{p-ampl}
\frac{\phi_\text{a}}{M} = \frac{\sqrt{2 \rho_\phi}}{m M} = \frac{2 \sqrt{\Omega_\text{\tiny DM}} H_0}{m} \sqrt{\frac{\rho_\phi}{\overline \rho_\phi}}\, \simeq \, 10^{- 30} \sqrt{\frac{\rho_\phi}{\overline \rho_\phi}}\, \frac{\text{meV}}{m}\, \lll \, 1 \, .
\end{equation}
Here, $\rho_\phi = m^2 \phi_\text{a}^2 / 2$ is the local energy density of scalaron dark matter, and $\overline \rho_\phi$ is its cosmological average today. The quantities $\Omega_\text{\tiny DM}$ and $H_0$ denote the present-day dark matter density parameter and the Hubble constant, respectively.

In the very early universe, the scalaron field is initially frozen due to Hubble friction and begins to oscillate when the Hubble damping scale $3H$ drops to the value of its mass $m$. For a radiation-dominated early universe, this condition reads
\begin{equation}\label{Hm}
3 H_\text{i} \equiv \sqrt{\frac{3 g_\text{i}}{5}} \frac{\pi T_\text{i}^2}{M} \simeq m \, ,
\end{equation}
where $g_\text{i} \approx 100$ denotes the number of relativistic degrees of freedom in thermal equilibrium at this epoch. It follows that $T_\text{i} \propto m^{1/2}$, and consequently $\left(\phi_\text{a} \right)_\text{i} / M \propto m^{-1/4}$.

Using the lower bound \eqref{mlow} on $m$, the estimate \eqref{p-ampl} yields $\left(\phi_\text{a} \right)_\text{i} / M \simeq 3.4 \times 10^{-7}$. This implies that the scalaron field remains extremely close to the minimum of its potential throughout cosmological evolution. Accordingly, only its mass term will be relevant in the following discussion.

\section{Effective scalaron--photon interaction in QED}
\label{sec:QED}

\subsection{General setup}

Before considering the effective scalaron--gauge field interactions in the full Standard Model, it is instructive to first examine the simpler case of quantum electrodynamics. Our starting point is QED with a single charged fermion $\psi$, described by the Lagrangian
\begin{equation} \label{spinor}
S \left[ e_a^\mu, \psi , A_\mu \right] = \int \left( \frac{ {\rm i}}{2}\, \overline \psi\! \stackrel{\leftrightarrow}{\slashed D}\! \psi - m_\psi \overline \psi \psi - \frac14 F_{\mu\nu} F^{\mu\nu} \right) \sqrt{-g}\, d^4 x \, , 
\end{equation}
where $\slashed D= \gamma^a e_a^\mu \left( \nabla_\mu + {\rm i} e A_\mu \right)$ is the covariant Dirac operator involving the electromagnetic field $A_\mu$ having field strength $F_{\mu\nu}$, $e$ is the fundamental electric charge, $e_a^\mu$ is the tetrad field, and $\nabla_\mu$ is the metric covariant derivative. We use the common notation $\stackrel{\leftrightarrow}{\slashed D} \ = \ \stackrel{\rightarrow}{\slashed D} - \stackrel{\leftarrow}{\slashed D}{\!\!}^\dagger$, where the arrow indicates the direction of action of this differential operator. We have put the kinetic part of the action for the fermion in a symmetric Hermitian form just for further convenience; all subsequent results remain the same starting from the usual non-symmetric action.

Performing the metric decomposition \eqref{om} in action \eqref{spinor}, we obtain the action in the form
\begin{equation} \label{LE}
\widetilde S \left[ \phi, \widetilde e_a^\mu, \psi, A_\mu \right] = \int \left( \frac{ {\rm i}}{2} e^{- {3 \phi}/{2 M}}\, \overline \psi\! \stackrel{\leftrightarrow}{\widetilde {\slashed D}}\! \psi - e^{- {2 \phi}/{M}} m_\psi \overline \psi \psi - \frac14 F_{\mu\nu} F^{\mu\nu} \right) \sqrt{- \widetilde g}\, d^4 x \, , 
\end{equation}
where $\widetilde {\slashed D} = \gamma^a \widetilde e_a^\mu \bigl( \widetilde \nabla_\mu + {\rm i} e A_\mu \bigr)$ is the covariant Dirac operator in the Einstein frame, in which $\widetilde e_a^\mu$ and $\widetilde \nabla_\mu$ are, respectively, the tetrad field and covariant derivative associated with the Einstein-frame metric $\widetilde g_{\mu\nu}$. Note that tensor indices in \eqref{LE} are raised and lowered with respect to the Einstein-frame metric.

To first order in $\phi/M$, the scalar field coupling in \eqref{LE} reads
\begin{equation} \label{Lint}
L_\text{int} = \frac{\phi}{2 M} T^\mu{}_\mu = \frac{\phi}{2 M} \left( - \frac{3 {\rm i}}{2} \overline \psi\! \stackrel{\leftrightarrow}{ {\slashed D}}\! \psi + 4 m_\psi \overline \psi \psi \right) \, .
\end{equation}
At one-loop level, the trace of the energy--momentum tensor in QED acquires an anomalous contribution \cite{Adler:1976zt}
\begin{equation} \label{Tan}
\left( T^\mu{}_\mu \right)_\text{anom} = \frac{\beta (e)}{2 e} F^{\mu\nu} F_{\mu\nu}  + \gamma_{m_\psi} m_\psi \overline \psi \psi \, ,
\end{equation}
where $\beta (e) = e^3 / 12 \pi^2$ is the beta-function for the electric charge $e$, and $\gamma_{m_\psi} = 3 e^2 / 8 \pi^2$ is the anomalous dimension of the fermion mass. As discussed in the Introduction, within the framework based on coupling to the Jordan-frame metric, this trace anomaly must be included in the effective scalaron--photon interaction [see equation \eqref{anom}]. In addition to the anomalous contribution, there is a purely diagrammatic one-loop contribution to the two-photon production amplitude. The full amplitude is given by the sum of these contributions, which we evaluate separately. For clarity, we emphasise that this paper deals exclusively with on-shell amplitudes.

\subsection{Trace-anomaly contribution}

In principle, one may simply use the first term in \eqref{Tan} as the trace-anomaly contribution to the scalaron coupling and then proceed to calculate the diagrammatic contribution from the classical trace in \eqref{Lint}. However, it is instructive to demonstrate that the one-loop contribution of the trace anomaly \eqref{Tan} to the effective scalaron--photon interaction can also be derived using the Fujikawa method \cite{Fujikawa:2004cx}. In this approach, the trace anomaly arises from the Jacobian of the functional measure under a fermion field transformation. We emphasise that the trace anomaly \eqref{Tan} can be obtained both from Feynman-diagram calculations using dimensional regularisation starting from the interaction $-\frac12 h_{\mu \nu} T^{\mu \nu}$ and then evaluating the trace, and from Fujikawa’s method; the two approaches yield identical results \cite{Kamada:2019pmx}.

One begins by introducing a diffeomorphism-invariant functional integration measure for the spinor semidensities $\psi' = (-g)^{1/4}\psi$ in the Jordan frame, which corresponds to quantising the spinor field in this frame. The action for the spinor semidensity $\psi'$ in the Jordan frame is
\begin{equation}\label{spinor'}
S' \left[ e_a^\mu, \psi^{\prime\,} \right] = \int \left( \frac{{\rm i}}{2}\, \overline \psi{}'\! \stackrel{\leftrightarrow}{\slashed D}\! \psi' - m_\psi \overline \psi{}' \psi' \right) d^4 x \, .
\end{equation}
Note that the flat-space limits of the spinor actions \eqref{spinor} and \eqref{spinor'} coincide. After the transformation $e_a^\mu = e^{\phi / 2 M}\,\widetilde e_a^\mu$ of the tetrad to the Einstein frame, action \eqref{spinor'} becomes
\begin{equation}\label{spinorE'}
S' \left[ e^{\phi / 2 M}\, \widetilde e_a^\mu, \psi^{\prime\,} \right] = \int \left( \frac{{\rm i}}{2} e^{\phi / 2 M}\, \overline \psi{}'\! \stackrel{\leftrightarrow}{\widetilde {\slashed D}}\! \psi' - m_\psi \overline \psi{}' \psi' \right) d^4 x \, .
\end{equation}

Now we eliminate the scalar $\phi$ from the kinetic term by a conformal transformation of the spinor:
\begin{equation}\label{psicon'}
\psi' = e^{- \phi / 4 M} \widetilde \psi{}' \, .
\end{equation}
After this transformation, Lagrangian \eqref{spinorE'} takes the form
\begin{equation} \label{spinorEt'}
\widetilde S' \left[ \phi, \widetilde e_a^\mu, \widetilde \psi{}' \right] = \int \left( \frac{{\rm i}}{2}\, \overline {\widetilde \psi{}'}\! \stackrel{\leftrightarrow}{\widetilde {\slashed D}}\! \widetilde \psi{}' - e^{- \phi / 2 M} m_\psi \overline {\widetilde \psi{}'} \widetilde \psi{}' \right) d^4 x \, . 
\end{equation}

Transformation \eqref{psicon'} introduces a non-trivial Jacobian in the fermionic functional integration measure over $\widetilde \psi'$ and $\overline{\widetilde \psi}{}'$ in the transformed action. Formally, the logarithm of this Jacobian can be written as
\begin{equation}\label{J-form}
\ln J = \ln \left| \Det \frac{{\cal D} \psi'}{{\cal D} \widetilde \psi{}'} \right|^{-2} = \frac12 \Tr \frac{\phi}{M} \, ,
\end{equation}
where $\Tr$ denotes the trace in fermion-field space. This trace requires regularisation, and its value depends on the choice of spinor basis. In Fujikawa's method, the trace in \eqref{J-form} is regularised in a gauge-invariant manner using a basis $\psi_n$ of eigenfunctions of the Dirac operator ${\rm i}\slashed D_\text{\tiny E}$ in Euclidean space that preserves the gauge symmetry. In this case, the regularised trace in the flat-space limit contains, in addition to an integral of an infinite constant multiplied by $\phi(x)$, a finite part describing an effective interaction between the scalar field and the gauge field (see \cite{Fujikawa:2004cx} and Appendix~\ref{app} for Euclidean space conventions):
\begin{equation} \label{canon}
\ln J_\text{reg} = \frac12 \left( \Tr \frac{\phi}{M} \right)_\text{reg} = {\rm i} \frac{e^2}{48 \pi^2} \int \frac{\phi}{M} F_{\mu\nu} F^{\mu\nu} d^4 x = {\rm i} \frac{\alpha}{12 \pi} \int \frac{\phi}{M} F_{\mu\nu} F^{\mu\nu} d^4 x \, ,
\end{equation}
where $\alpha = e^2 / 4 \pi$. This result reproduces the coupling of the scalar conformal mode to the electromagnetic field induced by the anomalous trace \eqref{Tan} of the energy--momentum tensor, which can be expressed as
\begin{equation}\label{fff-an-QED}
{\cal L}^\text{anom}_\text{\tiny QED} = \frac{\alpha}{16 \pi} F^\text{anom}_\text{\tiny QED} \frac{\phi}{M} F_{\mu\nu} F^{\mu\nu} \, , 
\end{equation} 
where $F^\text{anom}_\text{\tiny QED} = 4/3$.

\subsection{Diagrammatic contribution}

In addition to the Jacobian contribution reproducing the anomaly, there is a purely diagrammatic one-loop contribution to the two-photon production amplitude arising from the interaction term in \eqref{spinorEt'}. To first order in $\phi / M$, this contribution can be computed using either of the actions \eqref{LE} or \eqref{spinorE'}, yielding identical results and confirming \eqref{anom}. This reflects the field-redefinition equivalence theorem \cite{Chisholm:1961tha, Kamefuchi:1961sb, Divakaran:1963yxz, Arzt:1993gz, Cohen:2024fak}, according to which on-shell amplitudes are invariant under perturbative field redefinitions.  We now demonstrate this equivalence explicitly for the actions \eqref{spinorE'} and \eqref{spinorEt'}.

Lagrangian \eqref{spinorE'} can be written in the form
\begin{equation}\label{split}
L = \frac12 e^{\phi / 2 M} \left[ \overline \psi{}'  \bigl( {\rm i}\! \stackrel{\rightarrow}{\widetilde {\slashed D}} - m_\psi \bigr) \psi' - \overline \psi{}' \bigl( {\rm i}\! \stackrel{\leftarrow}{\widetilde {\slashed D}}{\!\!}^\dagger + m_\psi \bigr) \psi' \right] - \left( 1 - e^{\phi / 2 M} \right) m_\psi  \overline \psi{}' \psi' \, ,
\end{equation}
where we have simply rearranged the terms. One can show that the coupling arising from the first term (in square brackets) does not contribute to photon production through one-loop diagrams \cite{Shtanov:2025nue}.

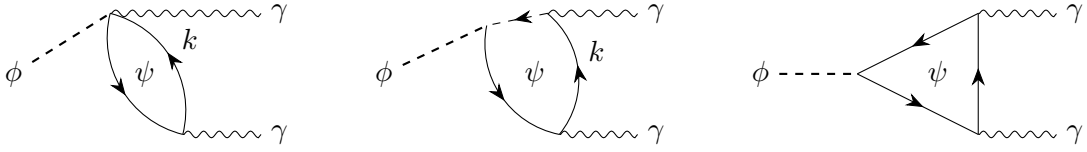
\begin{figure}
\centering
\begin{tikzpicture}[scale=1.6]
\begin{feynman}
\vertex (a) at (.3,0) {$\phi$};
\vertex (b) at (1.1,.5);
\vertex (c) at (1.7,-.5);
\vertex (e) at (2.5,-.5) {$\gamma$};
\vertex (f) at (2.5,.5) {$\gamma$};
\vertex (g) [label=0:$\psi$] at (1.22,0);
\diagram* { (a) -- [thick, scalar] (b), 
(b) -- [fermion, quarter right] (c) -- [fermion, quarter right, edge label'=$k$] (b), (b) -- [photon] (f), (c) -- [photon] (e)};
\end{feynman}
\end{tikzpicture} \qquad
\begin{tikzpicture}[scale=1.6]
\begin{feynman}
\vertex (a) at (.55,0) {$\phi$};
\vertex (b) at (1.4,.4);
\vertex (c) at (2,-.5);
\vertex (d) at (1.9,.5);
\vertex (e) at (2.8,-.5) {$\gamma$};
\vertex (f) at (2.8,.5) {$\gamma$};
\vertex (g) [label=0:$\psi$] at (1.63,0);
\diagram* { (a) -- [thick, scalar] (b), 
(b) -- [fermion, quarter right] (c) -- [fermion, quarter right, edge label'=$k$] (d) -- [fermion, dashed] (b), (c) -- [photon] (e), (d) -- [photon] (f)};
\end{feynman}
\end{tikzpicture} \qquad
\begin{tikzpicture}[scale=1.6]
\begin{feynman}
\vertex (a) at (.2,0) {$\phi$};
\vertex (b) at (1,0);
\vertex (c) at (2,-.5);
\vertex (d) at (2,.5);
\vertex (e) at (2.8,-.5) {$\gamma$};
\vertex (f) at (2.8,.5) {$\gamma$};
\vertex (g) [label=0:$\psi$] at (1.5,0);
\diagram* { (a) -- [thick, scalar] (b), 
(b) -- [fermion] (c) -- [fermion] (d) -- [fermion] (b), (c) -- [photon] (e), (d) -- [photon] (f)};
\end{feynman}
\end{tikzpicture}
\caption{The first two Feynman diagrams, arising from the first term (with square brackets) in \eqref{split}, are both quadratically divergent but their integrands cancel each other when the loop momentum $k$ is assigned consistently. The nonzero contribution comes from the third diagram arising from the last term in \eqref{split}. \label{fig:redef}}	
\end{figure}

Indeed, the first term in \eqref{split} (the expression in square brackets) contains the free Dirac equation, which is a so-called redundant operator. As a result, in the triangular diagram it effectively replaces one of the propagators by a constant, as indicated by the dashed edge in the second diagram of Fig.~\ref{fig:redef}. The contributions arising from the interaction vertex involving the scalar and the electromagnetic field, and from the free Dirac equation in \eqref{split}, are each quadratically divergent but cancel exactly when the loop momentum is treated consistently. These contributions are represented by the first two diagrams in Fig.~\ref{fig:redef}.

Moreover, each of the first two diagrams in Fig.~\ref{fig:redef}, while quadratically divergent in four dimensions, on shell depends only on one of the two photon momenta, $k_1$ or $k_2$. On the other hand, in any regularisation scheme that preserves gauge invariance, the on-shell amplitude must be proportional to $\left( k_1 \cdot k_2 \right) \left( \varepsilon_1 \cdot \varepsilon_2 \right) - \left( \varepsilon_1 \cdot k_2 \right) \left( \varepsilon_2 \cdot k_1 \right)$, where $\varepsilon_1$ and $\varepsilon_2$ are the photon polarisation vectors. This structure can only arise if each of these diagrams vanishes identically.

This can be verified by using dimensional regularisation. After a convenient shift of the loop momentum integration variable, each of these diagrams yields an expression proportional to $I_{\mu\nu} \varepsilon_1^\mu \varepsilon_2^\nu$, where
\begin{equation} \label{dim}
I_{\mu\nu} = \int d^4 k \left[ \frac{2 k_\mu k_\nu - g_{\mu\nu}  \bigl( k^2 - m_\psi^2 \bigr)}{\bigl( k^2 - m_\psi^2 + {\rm i} \epsilon \bigr)^2} \right] \, .
\end{equation}
This expression vanishes in dimensional regularisation. The same conclusion is obtained in Pauli--Villars regularisation.\footnote{Here we refer to the standard Pauli--Villars scheme, in which the propagators (or, equivalently, the full loop integrands) are modified by the introduction of regulator fields, while the interaction vertices remain unchanged. A different variant of Pauli--Villars regularisation will be discussed in section~\ref{sec:PV}.} Indeed, the subtraction terms introduced in this scheme have the same structure as the original diagram, differing only in their overall coefficients, signs, and fermion masses. The shift of the integration momentum is performed uniformly for all terms associated with a given diagram and is independent of the fermion masses. Consequently, the sum of all contributions yields a convergent integrand. The resulting expression may then be evaluated using dimensional regularisation, where each individual term is of the form \eqref{dim} and therefore vanishes. The total contribution is thus zero.

The remaining interaction terms in \eqref{split} and in \eqref{spinorEt'} are equivalent to linear order in $\phi/M$:
\begin{equation}
L_\text{int} = \frac{\phi}{2 M} m_\psi \overline \psi{}' \psi' \, .
\end{equation}
This gives rise to the third diagram in Fig.~\ref{fig:redef}. Its calculation involves a logarithmically divergent but conditionally convergent loop integral of the form \cite{Steinberger:1949wx}
\begin{equation}\label{log}
\int d^4 k \left[ \frac{4 k_\mu k_\nu - g_{\mu\nu} \left( k^2 - \Delta \right) }{\bigl( k^2 - \Delta + {\rm i} \epsilon \bigr)^3} \right] = 0 \, ,
\end{equation}
where the result is understood to be evaluated using dimensional regularisation or, equivalently, by invoking gauge invariance. The total result is therefore finite, yielding the well-known expression for the amplitude of a scalar decay into two photons mediated by a fermion loop \cite{Steinberger:1949wx}:
\begin{equation} \label{ampl}
{\cal M} = \frac{\alpha}{4 \pi M} F_\text{\tiny QED} (x) \bigl[ \left( \varepsilon_1 \cdot k_2 \right) \left( \varepsilon_2 \cdot k_1 \right) - \left( k_1 \cdot k_2 \right) \left( \varepsilon_1 \cdot \varepsilon_2 \right) \bigr] \, ,
\end{equation}
where the function $F_\text{\tiny QED} (x)$, with $x = 4 m_\psi^2 / m^2$, is given by
\begin{equation}
F_\text{\tiny QED} (x) = - 2 x \left[ 1 + (1 - x) f (x) \right] \, , \label{Ff}
\end{equation}
\begin{equation}\label{f}
f (x) = \left\{ 
\begin{array}{cl} 
\arcsin^2 x^{-1/2}  &\ \ \text{for} \ x \geq 1 \, , \\[5pt]
\dfrac14 \left( \pi + {\rm i} \ln \dfrac{1 + \sqrt{1 - x}}{1 - \sqrt{1 - x}} \right)^2  &\ \ \text{for} \ x < 1 \, .
\end{array} 
\right.
\end{equation}

Amplitude \eqref{ampl} can be reproduced by an effective interaction Lagrangian of the form
\begin{equation}\label{fff-diag-QED}
{\cal L}^\text{diag}_\text{\tiny QED} = \frac{\alpha}{16 \pi} F_\text{\tiny QED} (x) \frac{\phi}{M} F_{\mu\nu} F^{\mu\nu} \, . 
\end{equation}

\subsection{Total effective interaction}
\label{sec:total-QED}

The total effective interaction between the scalaron and electromagnetic field in QED is then given by the sum of \eqref{fff-an-QED} and \eqref{fff-diag-QED}:
\begin{equation}\label{fff-QED}
{\cal L}_\text{\tiny QED} = \frac{\alpha}{16 \pi} \left[ F_\text{\tiny QED} (x) + F^\text{anom}_\text{\tiny QED} \right] \frac{\phi}{M} F_{\mu\nu} F^{\mu\nu} \, , 
\end{equation} 
It determines the total on-shell amplitude of photon pair creation. This result was previously obtained in \cite{Kamada:2019pmx} by using dimensional regularisation of the energy--momentum tensor.

Note the limiting properties of the function $F_\text{\tiny QED} (x)$:
\begin{equation} \label{lim}
\lim_{x \to 0} F_\text{\tiny QED} (x) = 0 \, , \qquad \lim_{x \to \infty} F_\text{\tiny QED} (x) = - 4/3 = - F^\text{anom}_\text{\tiny QED} \, .
\end{equation}
Thus, in the massless-fermion limit, where QED is classically conformally invariant, the purely diagrammatic contribution to the effective scalar--electromagnetic interaction vanishes. In the opposite limit of an infinitely heavy fermion, the diagrammatic contribution yields a term equal in magnitude but opposite in sign to that induced by the trace anomaly, so that the total effective coupling in \eqref{fff-QED} cancels \cite{Kamada:2019pmx}.

This behavior reflects a general decoupling--anomaly matching structure: the diagrammatic contribution interpolates between vanishing in the conformal (massless) regime and cancelling the anomaly in the heavy-mass regime.

\subsection{Field implementation of Pauli--Villars regularisation}
\label{sec:PV}

It is noteworthy that the result \eqref{fff-QED} can also be reproduced within a particular gauge-invariant Pauli--Villars regularisation scheme, by reasoning similar to that employed in \cite{Kamada:2019pmx}. One of the transparent ways to implement Pauli--Villars regularisation with a single subtraction is to extend the Lagrangian by introducing two regulator fields: a bosonic spinor field $\zeta$ and a massive bosonic vector ghost field $B_\mu$, with the corresponding field strength $G_{\mu\nu}$ \cite{Schwartz:2014sze}. The vector ghost couples to the fermions in the same way as the gauge field $A_\mu$, but carries a negative-sign Lagrangian. The resulting extended action takes the form
\begin{align} \label{spinor-full}
S \left[ e_a^\mu, \psi, A_\mu, \zeta , B_\mu \right] = \int \left( \frac{{\rm i}}{2}\, \overline \psi\! \stackrel{\leftrightarrow}{\slashed D}\! \psi - m_\psi \overline \psi \psi - \frac14 F_{\mu\nu} F^{\mu\nu} \right) \sqrt{-g}\, d^4 x \nonumber \\ + \int \left( \frac{{\rm i}}{2}\, \overline \zeta\! \stackrel{\leftrightarrow}{\slashed D}\! \zeta - M_\zeta \overline \zeta \zeta + \frac14 G_{\mu\nu} G^{\mu\nu} - \frac12 M_\text{\tiny B}^2\, B_\mu B^\mu \right) \sqrt{-g}\, d^4 x \, . 
\end{align}

We now transform this action to the Einstein frame using the standard relation $e_a^\mu = e^{\phi / 2 M}\, \widetilde e_a^\mu$. Apart from the mass term of the vector ghost, the vector-field sector is conformally invariant and will be omitted. Written in terms of the spinor semidensities $\psi' = (-g)^{1/4}\psi$ and $\zeta' = (-g)^{1/4}\zeta$ realising the diffeomorphism-invariant functional integration measure, the spinor part of action \eqref{spinor-full} then becomes
\begin{align} \label{spinorE}
S \left[ e^{\phi / 2 M}\, \widetilde e_a^\mu, \psi', \zeta' \right] = \int \left( \frac{{\rm i}}{2} e^{\phi / 2 M}\, \overline \psi{}'\! \stackrel{\leftrightarrow}{\widetilde {\slashed D}}\! \psi' - m_\psi \overline \psi{}' \psi' \right) d^4 x \nonumber \\ + \int \left( \frac{{\rm i}}{2} e^{\phi / 2 M}\, \overline \zeta{}'\! \stackrel{\leftrightarrow}{\widetilde {\slashed D}}\! \zeta' - M_\zeta \overline \zeta{}' \zeta' \right) d^4 x \, . 
\end{align}

The scalar field $\phi$ can be removed from the kinetic terms of the spinors by the field redefinitions
\begin{equation}
\psi' = e^{- \phi / 4 M} \widetilde \psi{}' \, , \qquad \zeta' = e^{- \phi / 4 M} \widetilde \zeta{}' \, .
\end{equation}
Each of these transformations yields a formal Jacobian in the quantum functional integral. However, the two spinors have opposite statistics, so the corresponding Jacobians are mutually inverse (at least at one-loop order), and their product is therefore trivial. After the field redefinition, the spinor action becomes
\begin{align} \label{spinorEt}
\widetilde S \left[ \phi, \widetilde e_a^\mu, \widetilde \psi{}', \widetilde \zeta{}' \right] = \int \left( \frac{{\rm i}}{2}\, \overline {\widetilde \psi}{}'\! \stackrel{\leftrightarrow}{\widetilde {\slashed D}}\! \widetilde \psi{}' - e^{- \phi / 2 M} m_\psi \overline {\widetilde \psi}{}' \widetilde \psi{}' \right) d^4 x \nonumber \\ + \int \left( \frac{{\rm i}}{2}\, \overline {\widetilde \zeta}{}'\! \stackrel{\leftrightarrow}{\widetilde {\slashed D}}\! \widetilde \zeta{}'  - e^{- \phi / 2 M} M_\zeta \overline {\widetilde \zeta}{}'\, \widetilde \zeta{}' \right) d^4 x \, . 
\end{align}
The scalar field no longer couples to the fermion kinetic terms, while purely electromagnetic ultraviolet divergences in loop diagrams are regularised by the Pauli--Villars regulator fields. However, the ghost mass term still couples to the scalar field in the same way as the physical fermion mass term, albeit with its own mass parameter, and therefore contributes to photon production. Owing to the opposite statistics of the ghost field $\zeta$, the corresponding diagram in Fig.~\ref{fig:redef} with a ghost loop enters with an overall minus sign. As a result, we obtain
\begin{equation}
{\cal L}_\text{\tiny QED} = \frac{\alpha}{16 \pi} \left[ F_\text{\tiny QED} \left( 4 m_\psi^2 / m^2 \right) -  F_\text{\tiny QED} \left( 4 M_\zeta^2 / m^2 \right) \right] \frac{\phi}{M} F_{\mu\nu} F^{\mu\nu} \, . 
\end{equation} 
In the limit of $M_\zeta \to \infty$, by virtue of \eqref{lim}, this produces the result \eqref{fff-QED}. 

A key feature of the field-theoretic implementation of Pauli--Villars regularisation is that the coupling of the scalar $\phi$ to the ghost fermion $\zeta$ is proportional to its mass, whereas in the standard Pauli--Villars prescription only the loop propagators are modified, while the interaction vertices remain unchanged. It should also be noted that this approach ceases to be applicable when multiple Pauli--Villars subtractions are required to regularise loop integrals, since these cannot be implemented directly at the level of the Lagrangian.

\section{Effective scalaron--photon interactions in SM}
\label{sec:SM}

\subsection{Scalaron coupling to SM}

We assume that matter is described by the Standard Model minimally coupled to the metric in the Jordan frame. Most of the Standard Model action is classically conformally invariant, provided the matter fields are assigned the appropriate conformal transformation properties. The only sector that breaks classical conformal invariance is the Higgs sector, as well as any extension of the Standard Model neutrino sector containing Majorana mass terms.

The Higgs sector has the Lagrangian 
\begin{equation}\label{Sh}
L_\text{\tiny H} = g^{\mu\nu} \left( D_\mu \Phi \right)^\dagger D_\nu \Phi - \frac{\lambda}{4} \left( 2 \Phi^\dagger \Phi - v^2 \right)^2 \, .
\end{equation}
Here, $D_\mu$ is the gauge-covariant derivative involving the SU(2)$_L$ and U(1)$_Y$ electroweak gauge fields and acting on the Higgs doublet $\Phi$, while $v \approx 246\,\text{GeV}$ is the electroweak symmetry-breaking scale. In the Standard Model, the Higgs boson mass is given by $m_\text{\tiny H} = \sqrt{2 \lambda}\, v \approx 125\,\text{GeV}$, which corresponds to $\lambda \approx 0.13$.

After the conformal transformation \eqref{om}, the Higgs-sector Lagrangian takes the form (recalling the factor $\sqrt{-g}$ in the Lagrangian density)
\begin{equation}\label{Shn}
L_\text{\tiny H} = e^{- \phi / M} \widetilde g^{\mu\nu} \left( D_\mu \Phi \right)^\dagger D_\nu \Phi - \frac{\lambda}{4} e^{- 2 \phi / M} \left( 2 \Phi^\dagger \Phi - v^2 \right)^2 \, .
\end{equation}
The conformal transformation thus induces scalaron--Higgs field interactions suppressed by inverse powers of the Planck mass $M$. 

Similarly to the QED case discussed above, the scalaron couples directly to both the kinetic and mass terms of the fermionic fields in the Einstein frame. Since the kinetic terms of the Higgs and fermion fields contain gauge interactions, these couplings also induce interactions between the scalaron and the gauge fields. Such couplings can, however, be eliminated by performing the conformal transformations
\begin{equation} \label{Hcon}
\Phi = e^{\phi / 2 M} \widetilde \Phi \, , \qquad \psi = e^{3 \phi / 4 M} \widetilde \psi \, , 
\end{equation}
which bring the corresponding kinetic terms to their canonical Einstein-frame form. Similar transformations were considered in \cite{Burrage:2018dvt}. Under \eqref{Hcon}, the Lagrangian \eqref{Shn} becomes (all tensor contractions here and below are performed with the Einstein metric $\widetilde g^{\mu\nu}$)
\begin{equation}\label{Shdec}
L_\text{\tiny H} = \bigl( D_\mu \widetilde \Phi \bigr)^\dagger D^\mu \widetilde \Phi + \frac{1}{2 M} \partial_\mu \bigl( \widetilde \Phi^\dagger \widetilde \Phi \bigr) \partial^\mu \phi + \frac{1}{4 M^2} \widetilde \Phi^\dagger \widetilde \Phi\, \partial_\mu \phi\, \partial^\mu \phi - \frac{\lambda}{4} \left( 2 \widetilde \Phi^\dagger \widetilde \Phi - e^{- \phi / M} v^2 \right)^2 . 
\end{equation}
The scalaron now couples directly only to the Higgs sector and completely drops out of the remainder of the Standard Model Lagrangian, which assumes its canonical form, apart from possible Majorana mass terms.  

The scalaron and the Higgs field are slightly mixed in this field frame \cite{Shtanov:2021uif, Shtanov:2022xew, Shtanov:2025nue}. Adopting the unitary gauge for the Higgs doublet $\widetilde \Phi$ and expanding around its vacuum expectation value $v$, we write
\begin{equation}\label{Hshift}
\widetilde \Phi = \frac{1}{\sqrt{2}} \begin{pmatrix} 0 \\[2pt] \tilde h \end{pmatrix} = \frac{1}{\sqrt{2}} \begin{pmatrix} 0 \\ v + \varphi \end{pmatrix} \, , \end{equation}
where $\varphi$ is the shifted real-valued Higgs field. The quadratic part of the sum of Lagrangians \eqref{Sg3} and \eqref{Shdec} is then
\begin{equation}\label{L2}
L_2 = \frac12 \left( \partial \varphi \right)^2 + \frac12 \left( 1 + \frac{v^2}{4 M^2} \right) \left( \partial \phi \right)^2 + \frac{v}{2 M} \partial \varphi \partial \phi - \lambda v^2 \left( \varphi + \frac{v}{2 M} \phi \right)^2 - \frac12 m^2 \phi^2 \, .
\end{equation}
The field redefinition
\begin{equation}\label{shift}
\varphi = \chi - \frac{v}{2 M} \phi
\end{equation}
brings \eqref{L2} to the diagonal form
\begin{equation}
L_2 = \frac12 \left( \partial \chi \right)^2 + \frac12 \left( \partial \phi \right)^2 - \lambda v^2 \chi^2 - \frac12 m^2 \phi^2 \, .
\end{equation}
This parametrisation is particularly convenient for studying processes taking place well after the electroweak crossover, when the Higgs field has relaxed to its vacuum expectation value.

The original Jordan-frame Yukawa interaction $- \gamma h \overline \psi \psi$ between the Higgs field $h$ and a fermion $\psi$, where $\gamma$ is the Yukawa coupling constant, generates in the new frame the interaction
\begin{equation}\label{psicoup}
\frac{\gamma v}{2 M} \phi\, \overline{\widetilde \psi} \widetilde \psi = \frac{m_\psi}{2 M} \phi\, \overline{\widetilde \psi} \widetilde \psi \, ,
\end{equation}
where $m_\psi = \gamma v$ is the fermion mass. Analogous couplings are induced by Majorana mass terms, which transform to the Einstein frame as
\begin{equation}\label{Majorana}
M_\psi \overline {\psi^\text{c}_\text{\tiny R}} \psi^{}_\text{\tiny R} \sqrt{- g} \, \to \, e^{- \phi / 2 M} M_\psi \overline{\widetilde \psi^\text{c}_\text{\tiny R}} \widetilde \psi^{}_\text{\tiny R} \sqrt{- \widetilde g} \, .
\end{equation}

Likewise, the Higgs couplings to the electroweak gauge fields contained in \eqref{Shdec} generate, to first order in $\phi$, the interactions
\begin{equation}\label{gcoup}
- \frac{m_\text{\tiny W}^2}{M} \phi\, W_\mu^+ W^{-\mu} - \frac{m_\text{\tiny Z}^2}{2 M} \phi\, Z_\mu Z^\mu
\end{equation} 
between the scalaron and the vector bosons W$^\pm$ and Z$^0$. 

Interactions \eqref{psicoup} and \eqref{Majorana} also allow the scalaron to decay into fermion--antifermion pairs. If scalarons constitute the entirety of dark matter, a rough upper bound on the scalaron mass can be inferred from the observed 511~keV emission line from the Galactic Centre, which is generally interpreted as arising from electron--positron annihilation. Requiring that the positron flux from decaying scalaron dark matter remains consistent with the observed 511~keV signal yields the upper bound \cite{Cembranos:2008gj, Cembranos:2015svp}
\begin{equation}\label{mup}
m \lesssim 1.2~\text{MeV} \, .
\end{equation}

\subsection{Trace-anomaly contribution}

We first consider the trace-anomaly contribution to the effective scalaron--gauge field coupling. As in the QED case, this contribution is obtained either diagrammatically, starting from the interaction $-\frac12 h_{\mu \nu} T^{\mu \nu}$ and then evaluating the trace \cite{Coriano:2012nm}, or by using Fujikawa’s method \cite{Fujikawa:2004cx}, with both approaches yielding identical results. According to Fujikawa's approach, the conformal transformations \eqref{Hcon} remove the scalaron from the kinetic terms of the Higgs and fermion fields but generate a non-trivial Jacobian in the functional integral. This Jacobian describes anomalous interactions between the scalaron and the gauge bosons in the Einstein frame.

In curved spacetime, the functional integral that generates Feynman diagrams must be defined with a diffeo\-mor\-phism-invariant measure. In the Jordan frame, this measure is naturally expressed in terms of fields carrying conformal weight $1/2$:
\begin{equation} \label{primes}
\psi' = \left( - g \right)^{1/4} \psi \, , \qquad {\cal A}'_a = \left( - g \right)^{1/4} e_a^\mu {\cal A}_\mu \, , \qquad \Phi' = \left( - g \right)^{1/4} \Phi \, ,
\end{equation}
together with the corresponding gauge-fixing and ghost fields, which are likewise weighted by appropriate powers of the determinant $g$ of the Jordan-frame metric \cite{Fujikawa:2004cx}. Here, ${\cal A}_\mu$ denotes a generic gauge field, and $e_a^\mu$ is the tetrad. 

Consider an arbitrary transformation of these fields of the form
\begin{equation} \label{primet}
\psi' = \Omega^{- c_\text{F} /4}\, \widetilde \psi' \, , \qquad {\cal A}'_a = \Omega^{- c_\text{V} / 2} \widetilde {\cal A}'_a \, , \qquad \Phi' = \Omega^{- c_\text{S} / 2}\, \widetilde \Phi' \, ,
\end{equation}
with $\Omega$ given by \eqref{om}. For $c_\text{\tiny F} = c_\text{\tiny V} = c_\text{\tiny S} = 1$, this transformation reduces to the canonical conformal (Weyl) transformation. In that case, it generates a non-trivial Jacobian in the functional integral associated with the trace anomaly. The gauge-field-dependent part of the regularised Jacobian takes the form\footnote{There is ongoing controversy in the literature regarding the presence of parity-violating terms of the form $\tr {\cal F}^{\alpha\beta} {\cal F}^{\mu\nu} \epsilon_{\alpha \beta \mu \nu}$ in the chiral determinant (see \cite{Larue:2023tmu, Larue:2023qxw, Larue:2024zen, Bonora:2024imk}). In any case, such terms would cancel in scalaron interactions with photons and gluons, whose couplings to fermions are chirality symmetric.}\begin{equation}\label{jacob}
\ln J_\text{reg} = {\rm i} \frac{b_*^{} g_*^2}{64 \pi^2} \ln \Omega \sum_i {\cal F}^i_{\mu\nu} {\cal F}_i^{\mu\nu} \, ,
\end{equation}
where $g_*$ is the gauge coupling constant, and the coefficient $b_*$ determines the contribution of the corresponding gauge sector to the trace anomaly through the beta function of this coupling \cite{Fujikawa:2004cx}. Using the standard one-loop beta-function coefficients for the gauge groups and field content of the Standard Model \cite{Gross:1973ju, Srednicki:2010}, one finds
\begin{align}\label{bees}
b_\text{\tiny Y} = \frac{20}{9} c_\text{\tiny F} {\cal N} + \frac16 c_\text{\tiny S} \, , \quad b_\text{w} = -\frac{22}{3} c_\text{\tiny V} + \frac43 c_\text{\tiny F}\, {\cal N} + \frac16 c_\text{\tiny S} \, , \quad b_\text{s} = -11 c_\text{\tiny V} + \frac43 c_\text{\tiny F} \, {\cal N} \, ,
\end{align}
where ${\cal N} = 3$ is the number of fermion generations. The couplings $g_\text{\tiny Y}$, $g_\text{w}$, and $g_\text{s}$ correspond to the gauge groups U(1)$_\text{\tiny Y}$, SU(2)$_\text{\tiny L}$, and SU(3)$_\text{\tiny C}$, respectively. 

Starting from the Jordan-frame theory, we perform transformations \eqref{om} and \eqref{Hcon} to reach the Einstein frame, in which the Higgs and fermion fields have canonical kinetic terms. Consequently, transformation \eqref{primet} is conformal, corresponding to the coefficient values $c_\text{\tiny F} = c_\text{\tiny V} = c_\text{\tiny S} = 1$. Using $\ln \Omega = \phi / M$, we find that Jacobian \eqref{jacob} generates the following one-loop Lagrangian describing the anomalous couplings of the scalaron in the Standard Model with ${\cal N} = 3$:
\begin{equation} \label{anJ}
{\cal L}_\text{anom} = \frac{41 g_\text{\tiny Y}^2}{384 \pi^2} \frac{\phi}{M} B_{\mu\nu} B^{\mu\nu} -\frac{19 g_\text{w}^2}{384 \pi^2} \frac{\phi}{M} \sum_a W^a_{\mu\nu} W_a^{\mu\nu} - \frac{7 g_\text{s}^2}{64 \pi^2} \frac{\phi}{M}\, \sum_i G^i_{\mu\nu} G_i^{\mu\nu} \, .
\end{equation}
Here, $B_{\mu\nu}$ and $W_{\mu\nu}^a$, $a = 1, 2, 3$, are the field-strength tensors associated with the electroweak gauge groups U(1)$_\mathrm{Y}$ and SU(2)$_\mathrm{L}$, respectively, while $G^i_{\mu\nu}$, $i = 1, \ldots, 8$, is the gluon field-strength tensor.

To determine the anomalous coupling of the scalaron to the electromagnetic field strength $F_{\mu\nu}$, we use the relations 
\begin{align}
&B_{\mu\nu} = F_{\mu\nu} \cos \theta_\text{\tiny W} - Z_{\mu\nu} \sin \theta_\text{\tiny W} \, , \\
&W^3_{\mu\nu} = F_{\mu\nu} \sin \theta_\text{\tiny W} + Z_{\mu\nu} \cos \theta_\text{\tiny W} \, , \\
&g_\text{\tiny Y} \cos \theta_\text{\tiny W} = g_\text{w} \sin \theta_\text{\tiny W} = e \, ,
\end{align}
where $\theta_\text{\tiny W}$ is the weak mixing angle, $Z_{\mu\nu}$ is the field-strength tensor of the Z$^0$ boson, and $e$ is the fundamental electric charge. This gives
\begin{equation}\label{fff-an}
{\cal L}^\text{anom}_{\phi \gamma \gamma} = \frac{41 e^2}{384 \pi^2} \frac{\phi}{M} F_{\mu\nu} F^{\mu\nu} - \frac{19 e^2}{384 \pi^2} \frac{\phi}{M} F_{\mu\nu} F^{\mu\nu} = \frac{11 e^2}{192 \pi^2} \frac{\phi}{M} F_{\mu\nu} F^{\mu\nu} = \frac{\alpha}{16 \pi} F_\text{anom} \frac{\phi}{M} F_{\mu\nu} F^{\mu\nu} \, ,
\end{equation}
where $F_\text{anom} = 11/3$ is the trace anomaly coefficient in the Standard Model.

\subsection{Diagrammatic contribution}

To calculate the remaining diagrammatic contribution to the amplitude of the scalaron decay into photons, we proceed as in \cite{Shtanov:2025nue}, using reasoning similar to that of \cite{Goldberger:2007zk}. As shown in section~\ref{sec:SM}, the scalaron interactions \eqref{psicoup} and \eqref{gcoup}, to first order in $\phi$, match with those of the Higgs boson upon the substitution $\chi / v \to - \phi / 2 M$.

\begin{figure}
\centering
\begin{tikzpicture}[scale=1.57]
\begin{feynman}
\vertex (a) at (.2,0) {$\chi/\phi$};
\vertex (b) at (1,0);
\vertex (c) at (2,-.5);
\vertex (d) at (2,.5);
\vertex (e) at (2.8,-.5) {$\gamma$};
\vertex (f) at (2.8,.5) {$\gamma$};
\vertex (g) [label=0:$\psi$] at (1.5,0);
\diagram* { (a) -- [thick, scalar] (b), 
(b) -- [fermion] (c) -- [fermion] (d) -- [fermion] (b), (c) -- [photon] (e), (d) -- [photon] (f)};
\end{feynman}
\end{tikzpicture} \hspace{2pt} 
\begin{tikzpicture}[scale=1.57]
\begin{feynman}
\vertex (a) at (.2,0) {$\chi/\phi$};
\vertex (b) at (1,0);
\vertex (c) at (2,-.5);
\vertex (d) at (2,.5);
\vertex (e) at (2.8,-.5) {$\gamma$};
\vertex (f) at (2.8,.5) {$\gamma$};
\vertex (g) [label=0:$W$] at (1.45,0);
\diagram* { (a) -- [thick, scalar] (b), 
(b) -- [boson] (c) -- [boson] (d) -- [boson] (b), (c) -- [photon] (e), (d) -- [photon] (f)};
\end{feynman}
\end{tikzpicture} \hspace{2pt} 
\begin{tikzpicture}[scale=1.57]
\begin{feynman}
\vertex (a) at (.2,0) {$\chi/\phi$};
\vertex (b) at (1,0);
\vertex (c) at (2,0);
\vertex (d) at (2.8,-.5) {$\gamma$};
\vertex (e) at (2.8,.5) {$\gamma$};
\vertex (f) [label=0:$W$] at (1.3,0);
\diagram* { (a) -- [thick, scalar] (b), 
(b) -- [boson, half right] (c) -- [boson, half right] (b), (d) -- [photon] (c) -- [photon] (e)};
\end{feynman}
\end{tikzpicture}
\caption{Feynman diagrams in unitary gauge contributing to the decay of the Higgs boson $\chi$ or the scalaron $\phi$ into two photons $\gamma$.  \label{fig:decay}}	
\end{figure}
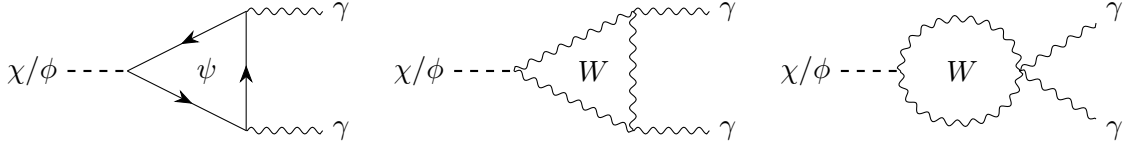

We recall that the effective one-loop interaction of the Higgs boson $\chi$ with photons in the Standard Model has the form
\begin{equation} \label{fff-H}
{\cal L}_{\text{\tiny H}\gamma \gamma} = - \frac{\alpha }{8 \pi} F \left( m_\text{\tiny H} \right) \frac{\chi}{v} F_{\mu\nu} F^{\mu\nu} \, . 
\end{equation}
The form factor $F (m)$ is given by \cite{Shifman:1979eb, Marciano:2011gm} \begin{equation}\label{F}
F (m) = F_\text{\tiny W} \left( x_\text{\tiny W} \right) + \sum_\psi  Q_\psi^2 F^{}_\text{\tiny F} \left( x^{}_\psi \right) \, .
\end{equation}
Here, the sum runs over all fermion fields, including the colour multiplicity of quarks; $Q_\psi$ denotes the fermion electric charge in units of the positron charge, and $x_i = 4 m_i^2 / m^2$. Furthermore,
\begin{align}
F_\text{\tiny F} (x) &\equiv F_\text{\tiny QED} (x) = - 2 x \left[ 1 + (1 - x) f (x) \right] \, ,\\[2pt]
F_\text{\tiny W} (x) &= 2 + 3 x + 3 x (2 - x) f (x) \, , 
\end{align}
where the function $f (x)$ is defined in \eqref{f}. This effective interaction describes the Higgs-boson decay into two photons, and the corresponding Feynman diagrams in unitary gauge are shown in figure~\ref{fig:decay}.

 Since the Higgs boson enters the diagrams of figure~\ref{fig:decay} only through a single vertex, the corresponding scalaron diagrams are evaluated simply by the substitution $\chi / v \to - \phi / 2 M$ in \eqref{fff-H},  which yields
\begin{equation}\label{fff-diag}
{\cal L}_{\phi \gamma \gamma}^\text{diag} = \frac{\alpha}{16 \pi} F (m) \frac{\phi}{M} F_{\mu\nu} F^{\mu\nu} \, , 
\end{equation}
where $F (m)$, now evaluated at the scalaron mass, is given by \eqref{F}. 

\subsection{Total effective interaction}
\label{sec:total-SM}

The total effective scalaron--photon interaction in the Standard Model is obtained as the sum of \eqref{fff-an} and \eqref{fff-diag}:
\begin{equation}\label{fff}
{\cal L}_{\phi \gamma \gamma} = \frac{\alpha}{16 \pi} \left[ F (m) + F_\text{anom} \right] \frac{\phi}{M} F_{\mu\nu} F^{\mu\nu} \, , 
\end{equation} 
The total decay width of the scalaron into two photons is equal to
\begin{equation}\label{Gamma}
\Gamma_{\phi\, \to\, \gamma \gamma} = \frac{\alpha^2 m^3}{2^{10} \pi^3 M^2} \left| F (m) + F_\text{anom} \right|^2 \, .
\end{equation}
In the Standard Model under consideration, one has $F_\text{anom} = 11/3$ according to \eqref{fff-an}. 

A similar result for the decay rate was obtained in \cite{Coriano:2012nm} for a dilaton coupled in the same manner to the trace of the energy--momentum tensor of the Standard Model, with the decay rate including only the contributions from the W$^\pm$ bosons and the top quark. Our result \eqref{Gamma} is derived for the specific case of scalaron in $f(R)$ gravity and includes contributions from all relevant SM fermions. It is directly applicable to the scalaron dark-matter scenario.

\begin{figure}[htp]
\begin{center}
\includegraphics[width=.7\textwidth]{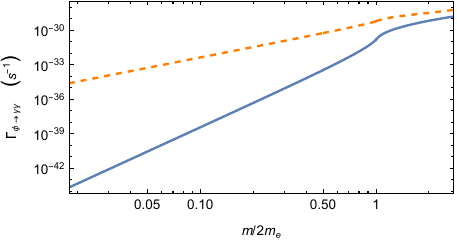}
\caption{Decay rate $\Gamma_{\phi\, \to\, \gamma\gamma}$ as a function of the scalaron mass $m$ on logarithmic scales for both axes. The solid blue curve shows the result of \eqref{Gamma} with the trace-anomaly contribution included ($F_\text{anom}=11/3$), while the dashed orange curve corresponds to $F_\text{anom}=0$. For $m<2m_e$, the decay rate scales as $\Gamma_{\phi\, \to\, \gamma\gamma}\propto m^7$ when the anomaly contribution is included, compared with $\Gamma_{\phi\, \to\, \gamma\gamma}\propto m^3$ when it is omitted. \label{fig:Gamma}}
\end{center}
\end{figure}

The decay rate \eqref{Gamma} with $F_\text{anom} = 11/3$ is shown in figure~\ref{fig:Gamma} as a solid blue curve as a function of the scalaron mass $m$. At the threshold $m = 2 m_e$, one finds
\begin{equation} \label{an-new}
\Gamma_{\phi\, \to\, \gamma\gamma} \approx 1.46 \times 10^{- 31}\, \text{s}^{-1} \approx \left( 6.85 \times 10^{30}\, \text{s} \right)^{-1} \, .
\end{equation}
For scalaron masses just below the electron--positron threshold, we have $F \approx -4$. For $m \ll 2 m_e$, one has $F + 11/3 \propto m^2$, so that $\Gamma_{\phi\, \to\, \gamma\gamma} \propto m^7$, exhibiting a rapid suppression at small $m$. This behaviour is analogous to the decoupling observed earlier in QED (see section~\ref{sec:total-QED}). More specifically, the effective scalaron--photon interaction is dominated by relatively light massive particles circulating in the loop, whereas heavy particles decouple. This contrasts with the effective Higgs--photon coupling, for which heavy charged particles yield dominant, non-decoupling contributions \cite{Goldberger:2007zk}.

\section{Different approaches in the literature}
\label{sec:frames}

In this paper, we treated the scalaron as an intrinsic part of the Jordan-frame metric to which the SM is minimally coupled. As explained in the Introduction, in this case, matrix elements are computed using the interaction $- \frac12 h_{\mu \nu} T^{\mu \nu}$, where $h_{\mu \nu}$ denotes the metric perturbation, and the trace is taken only at the final stage by setting $h_{\mu \nu} = - (\phi / M) \eta_{\mu \nu}$ according to \eqref{metric}. In practical terms, this means that the trace anomaly contributes to the effective scalaron--gauge field couplings, which we accounted for using the Fujikawa's approach. The anomalous contribution in \eqref{fff} and \eqref{Gamma} is then $F_\text{anom} = 11/3$.

As discussed in the Introduction, an alternative viewpoint is to treat the scalaron simply as an additional scalar field in the Einstein frame \cite{Watanabe:2010vy, Takeda:2014qma, Shtanov:2025nue}. Matrix elements are then computed using the classical-form interaction $(\phi / 2 M) \left( T^\mu{}_\mu \right)_\text{class}$, and the theory obeys the standard on-shell equivalence theorem associated with field redefinitions \cite{Shtanov:2025nue}. In the Einstein conformal frame for all fields, the scalaron is just mixed with the Higgs field, with mixing \eqref{shift} generating the coupling \eqref{fff} to electromagnetism. The anomalous interaction of the scalaron is absent in this approach, so that $F_\text{anom} = 0$ in \eqref{fff} and \eqref{Gamma}. The resulting decay rate into two photons \cite{Shtanov:2025nue} approximately matches that presented in \cite{Cembranos:2008gj} and is shown as the dashed orange curve in figure~\ref{fig:Gamma}. In the vicinity of $m = 2 m_e$, equation \eqref{Gamma} with $F_\text{anom} = 0$ gives
\begin{equation} \label{an-CS}
\Gamma_{\phi\, \to\, \gamma\gamma} \approx 5.2 \times 10^{- 30} \left( \frac{m}{\text{MeV}} \right)^3\, \text{s}^{-1} \approx \left[ 1.9 \times 10^{29} \left( \frac{\text{MeV}}{m} \right)^3\, \text{s}\, \right]^{-1} \, .
\end{equation}
This behaviour remains valid for smaller masses to within approximately $30\%$ accuracy \cite{Shtanov:2025nue}.

Should the scalar-field approach, which neglects the trace anomaly, be regarded as conceptually inconsistent? Perhaps not. Nevertheless, interpreting the scalaron as the conformal degree of freedom of the metric naturally incorporates the contribution of the trace anomaly and is therefore conceptually better motivated. From this perspective, the effective interaction and decay rate derived in section~\ref{sec:total-SM} provide a more complete and theoretically well-founded description.

At $m = 2 m_e$, the scalaron decay rate into photons predicted by \eqref{an-new} is approximately $36.5$ times smaller than the result \eqref{an-CS} obtained in the Einstein-frame approach, where the scalaron is treated as a scalar field that does not couple to the trace anomaly. Furthermore, as a consequence of the decoupling of heavy modes discussed in section~\ref{sec:total-SM}, the new decay rate scales as $\Gamma_{\phi\, \to\, \gamma \gamma}\propto m^7$ at low masses, rather than $\Gamma_{\phi\, \to\, \gamma \gamma}\propto m^3$ as in \eqref{an-CS} (see figure~\ref{fig:Gamma}).

We recall that the electromagnetic signal from scalaron dark matter would manifest itself as a monochromatic photon line produced by the decay process $\phi \to \gamma \gamma$ \cite{Cembranos:2008gj}. The substantially reduced decay rate predicted by \eqref{an-new}, together with its stronger suppression at low masses, implies a significantly weaker photon-line signal, thereby substantially reducing the prospects for detecting scalaron dark matter through its purely electromagnetic decay channel.

\section{Conclusions}
\label{sec:summary}

In this work, we have revisited the effective coupling of the scalaron to gauge fields in $f(R)$ gravity minimally coupled to the Standard Model, focusing on the scalaron decay into two photons. Our analysis was carried out within the Jordan-frame metric viewpoint \cite{Dolgov:1980kp, Dolgov:1981nw, Coriano:2012nm, Gorbunov:2012ns, Kamada:2019pmx}, in which the scalaron is treated as the conformal degree of freedom of the metric and quantum amplitudes are derived from its coupling to the anomalous energy--momentum tensor. Within this framework, the trace anomaly contributes explicitly to the scalaron--gauge boson interaction.

The anomaly-induced contribution to the effective scalaron--gauge boson interaction can be derived either diagrammatically, treating the full metric as the fundamental field, or within the scalaron framework using Fujikawa's method. The two approaches lead to identical results. In Fujikawa's approach, the origin of the anomaly-induced contribution is associated with the transformation of matter fields between the Jordan and Einstein frames, and the relevant Jacobian is related to the beta-functions of the corresponding gauge coupling constants. We first illustrated the procedure in QED and subsequently extended the analysis to the full Standard Model. The resulting effective scalaron--photon interaction reproduces the structure previously obtained in the diagrammatic approach taking into account the trace anomaly \cite{Coriano:2012nm, Kamada:2019pmx} and leads to a decay rate that differs from that obtained using only the interaction $(\phi/2M) \left( T^\mu{}_\mu \right)_\text{class}$ with the classical form of the energy--momentum tensor \cite{Watanabe:2010vy, Takeda:2014qma, Shtanov:2025nue}. In particular, in the limit $m \ll m_i$, where $m$ is the scalaron mass and $m_i$ denote the masses of particles running in the loop, the diagrammatic contribution from the interaction $(\phi/2M) \left( T^\mu{}_\mu \right)_\text{class}$ exactly cancels the anomaly-induced term, indicating the decoupling of heavy particles \cite{Kamada:2019pmx}. Consequently, the scalaron decay rate into photons is substantially more suppressed at low scalaron masses.

Our results clarify the origin of the discrepancy in the literature concerning the scalaron coupling to massless gauge fields. This difference does not originate from the classical $f(R)$ theory itself, whose Jordan- and Einstein-frame formulations are related by a field redefinition, but rather from the treatment of quantum effects associated with conformal transformations. From the Jordan-frame metric viewpoint, the trace anomaly appears explicitly, whereas in the Einstein-frame approach it is sometimes not introduced as a separate contribution. The two procedures therefore correspond to distinct prescriptions for implementing quantum corrections in $f(R)$ gravity and, as a consequence, yield different effective scalaron--gauge field interactions for processes involving massless gauge bosons.

From a phenomenological perspective, the scalaron decay rate into photons is of particular interest for dark-matter searches based on monochromatic photon signals. The results obtained here provide the corresponding decay rate within the Jordan-frame metric viewpoint and may be used in future studies of scalaron dark matter and its observational signatures. More generally, our analysis highlights the nontrivial role of trace anomaly and frame transformations in modified gravity theories.

\acknowledgments

This research was funded by the National Research Foundation of Ukraine under project 2023.03/0149.

\appendix

\section{Euclidean space conventions}
\label{app}

Proceeding to Euclidean curved space, one performs the Wick rotation by transforming the tetrads and the $\gamma^0$ matrix,
\begin{equation}
e_0^\mu = {\rm i} \left( e_{0}^{\mu} \right)_\text{\tiny E} \, , \qquad e^0_\mu = - {\rm i} \left( e^{0}_{\mu} \right)_\text{\tiny E}\, , \qquad \gamma^0 = - {\rm i} \left( \gamma^0 \right)_\text{\tiny E} \, ,
\end{equation}
after which the Euclidean quantities are taken to be real. One finds that $F^{\mu\nu} F_{\mu\nu} \det {e^a_\mu} = - {\rm i} \left( F^{\mu\nu} F_{\mu\nu} \det {e^a_\mu} \right)_\text{\tiny E}$. The Dirac operator $\slashed D$ is form-invariant under this continuation.

Regularised traces are evaluated in Euclidean space. For the gauge-field-dependent part in QED, using a gauge-symmetry-preserving basis $\psi_n$ of the Dirac operator ${\rm i}\slashed D_\text{\tiny E}$, one has \cite{Fujikawa:2004cx}
\begin{align} 
\left(\Tr \phi \right)_\text{reg} &= \lim_{\Lambda \to \infty} \Tr \phi \, e^{- \slashed D_\text{E}^2 / \Lambda^2} = \lim_{\Lambda \to \infty} \int d^4 x\, \phi (x) \sum_n \overline \psi_n (x) e^{- \slashed D_\text{E}^2 / \Lambda^2} \psi_n (x) \nonumber \\ &= \frac{e^2}{24 \pi^2} \int \phi \left( F^{\mu\nu} F_{\mu\nu} \det {e^a_\mu} \right)_\text{\tiny E} d^4 x = {\rm i} \frac{e^2}{24 \pi^2} \int \phi F^{\mu\nu} F_{\mu\nu} \det {e^a_\mu}\, d^4 x \, .
\end{align}

\bibliographystyle{JHEP}
\bibliography{scalaron}

\providecommand{\href}[2]{#2}\begingroup\raggedright\begin{thebibliography}{10}

\bibitem{Starobinsky:1980te}
A.A.~Starobinsky, \emph{A new type of isotropic cosmological models without
  singularity}, \href{https://doi.org/10.1016/0370-2693(80)90670-X}{\emph{Phys.
  Lett. B} {\bfseries 91} (1980) 99}.

\bibitem{Vilenkin:1985md}
A.~Vilenkin, \emph{Classical and quantum cosmology of the {S}tarobinsky
  inflationary model},
  \href{https://doi.org/10.1103/PhysRevD.32.2511}{\emph{Phys. Rev. D}
  {\bfseries 32} (1985) 2511}.

\bibitem{Sotiriou:2008rp}
T.P.~Sotiriou and V.~Faraoni, \emph{{$f(R)$} theories of gravity},
  \href{https://doi.org/10.1103/RevModPhys.82.451}{\emph{Rev. Mod. Phys.}
  {\bfseries 82} (2010) 451} [\href{https://arxiv.org/abs/0805.1726}{{\ttfamily
  0805.1726}}].

\bibitem{DeFelice:2010aj}
A.~De~Felice and S.~Tsujikawa, \emph{{$f(R)$} theories},
  \href{https://doi.org/10.12942/lrr-2010-3}{\emph{Living Rev. Rel.} {\bfseries
  13} (2010) 3} [\href{https://arxiv.org/abs/1002.4928}{{\ttfamily
  1002.4928}}].

\bibitem{Nojiri:2010wj}
S.~Nojiri and S.D.~Odintsov, \emph{Unified cosmic history in modified gravity:
  {F}rom {$F(R)$} theory to {L}orentz non-invariant models},
  \href{https://doi.org/10.1016/j.physrep.2011.04.001}{\emph{Phys. Rept.}
  {\bfseries 505} (2011) 59} [\href{https://arxiv.org/abs/1011.0544}{{\ttfamily
  1011.0544}}].

\bibitem{Capozziello:2006uv}
S.~Capozziello, V.F.~Cardone and A.~Troisi, \emph{Dark energy and dark matter
  as curvature effects},
  \href{https://doi.org/10.1088/1475-7516/2006/08/001}{\emph{JCAP} {\bfseries
  08} (2006) 001} [\href{https://arxiv.org/abs/astro-ph/0602349}{{\ttfamily
  astro-ph/0602349}}].

\bibitem{Nojiri:2008nt}
S.~Nojiri and S.D.~Odintsov, \emph{Dark energy, inflation and dark matter from
  modified {$F(R)$} gravity}, {\emph{TSPU Bulletin} {\bfseries N8(110)} (2011)
  7} [\href{https://arxiv.org/abs/0807.0685}{{\ttfamily 0807.0685}}].

\bibitem{Cembranos:2008gj}
J.A.R.~Cembranos, \emph{Dark matter from {$R^2$} gravity},
  \href{https://doi.org/10.1103/PhysRevLett.102.141301}{\emph{Phys. Rev. Lett.}
  {\bfseries 102} (2009) 141301}
  [\href{https://arxiv.org/abs/0809.1653}{{\ttfamily 0809.1653}}].

\bibitem{Cembranos:2015svp}
J.A.R.~Cembranos, \emph{Modified gravity and dark matter},
  \href{https://doi.org/10.1088/1742-6596/718/3/032004}{\emph{J. Phys. Conf.
  Ser.} {\bfseries 718} (2016) 032004}
  [\href{https://arxiv.org/abs/1512.08752}{{\ttfamily 1512.08752}}].

\bibitem{Corda:2011aa}
C.~Corda, H.J.~Mosquera~Cuesta and R.~Lorduy~G{\'o}mez, \emph{High-energy
  scalarons in {$R^{2}$} gravity as a model for {D}ark {M}atter in galaxies},
  \href{https://doi.org/10.1016/j.astropartphys.2011.08.009}{\emph{Astropart.
  Phys.} {\bfseries 35} (2012) 362}
  [\href{https://arxiv.org/abs/1105.0147}{{\ttfamily 1105.0147}}].

\bibitem{Katsuragawa:2016yir}
T.~Katsuragawa and S.~Matsuzaki, \emph{Dark matter in modified gravity?},
  \href{https://doi.org/10.1103/PhysRevD.95.044040}{\emph{Phys. Rev. D}
  {\bfseries 95} (2017) 044040}
  [\href{https://arxiv.org/abs/1610.01016}{{\ttfamily 1610.01016}}].

\bibitem{Katsuragawa:2017wge}
T.~Katsuragawa and S.~Matsuzaki, \emph{Cosmic history of chameleonic dark
  matter in {$F(R)$} gravity},
  \href{https://doi.org/10.1103/PhysRevD.97.064037}{\emph{Phys. Rev. D}
  {\bfseries 97} (2018) 064037}
  [\href{https://arxiv.org/abs/1708.08702}{{\ttfamily 1708.08702}}].

\bibitem{Yadav:2018llv}
B.K.~Yadav and M.M.~Verma, \emph{Dark matter as scalaron in {$f(R)$} gravity
  models}, \href{https://doi.org/10.1088/1475-7516/2019/10/052}{\emph{JCAP}
  {\bfseries 10} (2019) 052}
  [\href{https://arxiv.org/abs/1811.03964}{{\ttfamily 1811.03964}}].

\bibitem{Parbin:2020bpp}
N.~Parbin and U.D.~Goswami, \emph{Scalarons mimicking dark matter in the
  {H}u-{S}awicki model of {$f(R)$} gravity},
  \href{https://doi.org/10.1142/S0217732321502655}{\emph{Mod. Phys. Lett. A}
  {\bfseries 36} (2021) 2150265}
  [\href{https://arxiv.org/abs/2007.07480}{{\ttfamily 2007.07480}}].

\bibitem{KumarSharma:2022qdf}
V.~Kumar~Sharma and M.M.~Verma, \emph{Unified {$f(R)$} gravity at local
  scales}, \href{https://doi.org/10.1140/epjc/s10052-022-10329-6}{\emph{Eur.
  Phys. J. C} {\bfseries 82} (2022) 400}
  [\href{https://arxiv.org/abs/2201.01058}{{\ttfamily 2201.01058}}].

\bibitem{Shtanov:2021uif}
Y.~Shtanov, \emph{Light scalaron as dark matter},
  \href{https://doi.org/10.1016/j.physletb.2021.136469}{\emph{Phys. Lett. B}
  {\bfseries 820} (2021) 136469}
  [\href{https://arxiv.org/abs/2105.02662}{{\ttfamily 2105.02662}}].

\bibitem{Shtanov:2022xew}
Y.~Shtanov, \emph{Initial conditions for the scalaron dark matter},
  \href{https://doi.org/10.1088/1475-7516/2022/10/079}{\emph{JCAP} {\bfseries
  10} (2022) 079} [\href{https://arxiv.org/abs/2207.00267}{{\ttfamily
  2207.00267}}].

\bibitem{Shtanov:2024nmf}
Y.~Shtanov, \emph{Scalaron dark matter and the thermal history of the
  universe}, \href{https://doi.org/10.1088/1475-7516/2024/12/028}{\emph{JCAP}
  {\bfseries 12} (2024) 028}
  [\href{https://arxiv.org/abs/2409.05027}{{\ttfamily 2409.05027}}].

\bibitem{Shtanov:2025nue}
Y.~Shtanov and Y.~Sheiko, \emph{Interactions of the scalaron dark matter in {$f
  (R)$} gravity},
  \href{https://doi.org/10.1088/1475-7516/2025/07/033}{\emph{JCAP} {\bfseries
  07} (2025) 033} [\href{https://arxiv.org/abs/2505.00324}{{\ttfamily
  2505.00324}}].

\bibitem{Watanabe:2010vy}
Y.~Watanabe, \emph{Rate of gravitational inflaton decay via gauge trace
  anomaly}, \href{https://doi.org/10.1103/PhysRevD.83.043511}{\emph{Phys. Rev.
  D} {\bfseries 83} (2011) 043511}
  [\href{https://arxiv.org/abs/1011.3348}{{\ttfamily 1011.3348}}].

\bibitem{Takeda:2014qma}
N.~Takeda and Y.~Watanabe, \emph{No quasistable scalaron lump forms after
  {$R^2$} inflation},
  \href{https://doi.org/10.1103/PhysRevD.90.023519}{\emph{Phys. Rev. D}
  {\bfseries 90} (2014) 023519}
  [\href{https://arxiv.org/abs/1405.3830}{{\ttfamily 1405.3830}}].

\bibitem{Dolgov:1980kp}
A.D.~Dolgov, \emph{Massless particle production by conformal plane gravitation
  field. ({I}n {R}ussian)}, {\emph{Pisma Zh. Eksp. Teor. Fiz.} {\bfseries 32}
  (1980) 673}.

\bibitem{Dolgov:1981nw}
A.D.~Dolgov, \emph{Conformal anomaly and the production of massless particles
  by a conformally flat metric}, {\emph{Sov. Phys. JETP} {\bfseries 54} (1981)
  223}.

\bibitem{Giannotti:2008cv}
M.~Giannotti and E.~Mottola, \emph{Trace anomaly and massless scalar degrees of
  freedom in gravity},
  \href{https://doi.org/10.1103/PhysRevD.79.045014}{\emph{Phys. Rev. D}
  {\bfseries 79} (2009) 045014}
  [\href{https://arxiv.org/abs/0812.0351}{{\ttfamily 0812.0351}}].

\bibitem{Armillis:2009pq}
R.~Armillis, C.~Corian\`{o} and L.~Delle~Rose, \emph{Conformal anomalies and
  the gravitational effective action: {T}he {$TJJ$} correlator for a {D}irac
  fermion}, \href{https://doi.org/10.1103/PhysRevD.81.085001}{\emph{Phys. Rev.
  D} {\bfseries 81} (2010) 085001}
  [\href{https://arxiv.org/abs/0910.3381}{{\ttfamily 0910.3381}}].

\bibitem{Coriano:2012nm}
C.~Corian\`{o}, L.~Delle~Rose, A.~Quintavalle and M.~Serino, \emph{Dilaton
  interactions and the anomalous breaking of scale invariance of the {S}tandard
  {M}odel}, \href{https://doi.org/10.1007/JHEP06(2013)077}{\emph{JHEP}
  {\bfseries 06} (2013) 077} [\href{https://arxiv.org/abs/1206.0590}{{\ttfamily
  1206.0590}}].

\bibitem{Gorbunov:2012ns}
D.~Gorbunov and A.~Tokareva, \emph{{$R^2$}-inflation with conformal {SM H}iggs
  field}, \href{https://doi.org/10.1088/1475-7516/2013/12/021}{\emph{JCAP}
  {\bfseries 12} (2013) 021} [\href{https://arxiv.org/abs/1212.4466}{{\ttfamily
  1212.4466}}].

\bibitem{Kamada:2019pmx}
A.~Kamada, \emph{On scalaron decay via the trace of energy-momentum tensor},
  \href{https://doi.org/10.1007/JHEP07(2019)172}{\emph{JHEP} {\bfseries 07}
  (2019) 172} [\href{https://arxiv.org/abs/1902.05209}{{\ttfamily
  1902.05209}}].

\bibitem{Planck:2018jri}
{\scshape Planck} collaboration, \emph{Planck 2018 results. {X}. {C}onstraints
  on inflation},
  \href{https://doi.org/10.1051/0004-6361/201833887}{\emph{Astron. Astrophys.}
  {\bfseries 641} (2020) A10}
  [\href{https://arxiv.org/abs/1807.06211}{{\ttfamily 1807.06211}}].

\bibitem{Wald:1984}
R.M.~Wald, \emph{General Relativity}, University of Chicago Press, Chicago
  (1984),
  \href{https://doi.org/10.7208/chicago/9780226870373.001.0001}{10.7208/chicago/9780226870373.001.0001}.

\bibitem{Shtanov:2022pdx}
Y.~Shtanov, V.~Sahni and S.S.~Mishra, \emph{Tabletop potentials for inflation
  from {$f(R)$} gravity},
  \href{https://doi.org/10.1088/1475-7516/2023/03/023}{\emph{JCAP} {\bfseries
  03} (2023) 023} [\href{https://arxiv.org/abs/2210.01828}{{\ttfamily
  2210.01828}}].

\bibitem{Dicke:1961gz}
R.H.~Dicke, \emph{Mach's principle and invariance under transformation of
  units}, \href{https://doi.org/10.1103/PhysRev.125.2163}{\emph{Phys. Rev.}
  {\bfseries 125} (1962) 2163}.

\bibitem{Faraoni:2006fx}
V.~Faraoni and S.~Nadeau, \emph{({P}seudo)issue of the conformal frame
  revisited}, \href{https://doi.org/10.1103/PhysRevD.75.023501}{\emph{Phys.
  Rev. D} {\bfseries 75} (2007) 023501}
  [\href{https://arxiv.org/abs/gr-qc/0612075}{{\ttfamily gr-qc/0612075}}].

\bibitem{Shtanov:2022wpr}
Y.~Shtanov, \emph{On the conformal frames in {$f(R)$} gravity},
  \href{https://doi.org/10.3390/universe8020069}{\emph{Universe} {\bfseries 8}
  (2022) 69} [\href{https://arxiv.org/abs/2202.00818}{{\ttfamily 2202.00818}}].

\bibitem{Stelle:1977ry}
K.S.~Stelle, \emph{Classical gravity with higher derivatives},
  \href{https://doi.org/10.1007/BF00760427}{\emph{Gen. Rel. Grav.} {\bfseries
  9} (1978) 353}.

\bibitem{Lee:2020zjt}
J.G.~Lee, E.G.~Adelberger, T.S.~Cook, S.M.~Fleischer and B.R.~Heckel, \emph{New
  test of the gravitational $1/r^2$ law at separations down to 52 $\mu$m},
  \href{https://doi.org/10.1103/PhysRevLett.124.101101}{\emph{Phys. Rev. Lett.}
  {\bfseries 124} (2020) 101101}
  [\href{https://arxiv.org/abs/2002.11761}{{\ttfamily 2002.11761}}].

\bibitem{Adler:1976zt}
S.L.~Adler, J.C.~Collins and A.~Duncan, \emph{Energy-momentum-tensor trace
  anomaly in spin-1/2 quantum electrodynamics},
  \href{https://doi.org/10.1103/PhysRevD.15.1712}{\emph{Phys. Rev. D}
  {\bfseries 15} (1977) 1712}.

\bibitem{Fujikawa:2004cx}
K.~Fujikawa and H.~Suzuki, \emph{Path Integrals and Quantum Anomalies},
  Clarendon Press, Oxford, UK (2004),
  \href{https://doi.org/10.1093/acprof:oso/9780198529132.001.0001}{10.1093/acprof:oso/9780198529132.001.0001}.

\bibitem{Chisholm:1961tha}
J.S.R.~Chisholm, \emph{Change of variables in quantum field theories},
  \href{https://doi.org/10.1016/0029-5582(61)90106-7}{\emph{Nucl. Phys.}
  {\bfseries 26} (1961) 469}.

\bibitem{Kamefuchi:1961sb}
S.~Kamefuchi, L.~O'Raifeartaigh and A.~Salam, \emph{Change of variables and
  equivalence theorems in quantum field theories},
  \href{https://doi.org/10.1016/0029-5582(61)90056-6}{\emph{Nucl. Phys.}
  {\bfseries 28} (1961) 529}.

\bibitem{Divakaran:1963yxz}
P.P.~Divakaran, \emph{Equivalence theorems and point transformations in field
  theory}, \href{https://doi.org/10.1016/0029-5582(63)90731-4}{\emph{Nucl.
  Phys.} {\bfseries 42} (1963) 235}.

\bibitem{Arzt:1993gz}
C.~Arzt, \emph{Reduced effective lagrangians},
  \href{https://doi.org/10.1016/0370-2693(94)01419-D}{\emph{Phys. Lett. B}
  {\bfseries 342} (1995) 189}
  [\href{https://arxiv.org/abs/hep-ph/9304230}{{\ttfamily hep-ph/9304230}}].

\bibitem{Cohen:2024fak}
T.~Cohen, M.~Forslund and A.~Helset, \emph{Field redefinitions can be
  nonlocal},  \href{https://arxiv.org/abs/2412.12247}{{\ttfamily 2412.12247}}.

\bibitem{Steinberger:1949wx}
J.~Steinberger, \emph{On the use of subtraction fields and the lifetimes of
  some types of meson decay},
  \href{https://doi.org/10.1103/PhysRev.76.1180}{\emph{Phys. Rev.} {\bfseries
  76} (1949) 1180}.

\bibitem{Schwartz:2014sze}
M.D.~Schwartz, \emph{Quantum Field Theory and the Standard Model}, Cambridge
  University Press, Cambridge (2014),
  \href{https://doi.org/10.1017/9781139540940}{10.1017/9781139540940}.

\bibitem{Burrage:2018dvt}
C.~Burrage, E.J.~Copeland, P.~Millington and M.~Spannowsky, \emph{Fifth forces,
  {H}iggs portals and broken scale invariance},
  \href{https://doi.org/10.1088/1475-7516/2018/11/036}{\emph{JCAP} {\bfseries
  11} (2018) 036} [\href{https://arxiv.org/abs/1804.07180}{{\ttfamily
  1804.07180}}].

\bibitem{Larue:2023tmu}
R.~Larue, J.~Quevillon and R.~Zwicky, \emph{Trace anomaly of {W}eyl fermions
  via the path integral},
  \href{https://doi.org/10.1007/JHEP12(2023)064}{\emph{JHEP} {\bfseries 12}
  (2023) 064} [\href{https://arxiv.org/abs/2309.08670}{{\ttfamily
  2309.08670}}].

\bibitem{Larue:2023qxw}
R.~Larue, J.~Quevillon and R.~Zwicky, \emph{Gravity-gauge anomaly constraints
  on the energy-momentum tensor},
  \href{https://doi.org/10.1007/JHEP05(2024)307}{\emph{JHEP} {\bfseries 05}
  (2024) 307} [\href{https://arxiv.org/abs/2312.13222}{{\ttfamily
  2312.13222}}].

\bibitem{Larue:2024zen}
R.~Larue, J.~Quevillon and R.~Zwicky, \emph{Model-independent results on parity
  violation in the trace anomaly},  in \emph{{17th Marcel Grossmann Meeting}:
  {On Recent Developments in Theoretical and Experimental General Relativity,
  Gravitation, and Relativistic Field Theories}}, 11, 2024
  [\href{https://arxiv.org/abs/2411.00571}{{\ttfamily 2411.00571}}].

\bibitem{Bonora:2024imk}
L.~Bonora and S.G.~Giaccari, \emph{Something anomalies can tell about
  {S}tandard {M}odel and gravity},
  \href{https://doi.org/10.3390/sym17020273}{\emph{Symmetry} {\bfseries 17}
  (2025) 273} [\href{https://arxiv.org/abs/2412.07470}{{\ttfamily
  2412.07470}}].

\bibitem{Gross:1973ju}
D.J.~Gross and F.~Wilczek, \emph{Asymptotically free gauge theories. {I}},
  \href{https://doi.org/10.1103/PhysRevD.8.3633}{\emph{Phys. Rev. D} {\bfseries
  8} (1973) 3633}.

\bibitem{Srednicki:2010}
M.~Srednicki, \emph{Quantum Field Theory}, Cambridge University Press,
  Cambridge, fourth~ed. (2010).

\bibitem{Goldberger:2007zk}
W.D.~Goldberger, B.~Grinstein and W.~Skiba, \emph{Distinguishing the {H}iggs
  boson from the dilaton at the {L}arge {H}adron {C}ollider},
  \href{https://doi.org/10.1103/PhysRevLett.100.111802}{\emph{Phys. Rev. Lett.}
  {\bfseries 100} (2008) 111802}
  [\href{https://arxiv.org/abs/0708.1463}{{\ttfamily 0708.1463}}].

\bibitem{Shifman:1979eb}
M.A.~Shifman, A.I.~Vainshtein, M.B.~Voloshin and V.I.~Zakharov,
  \emph{Low-energy theorems for {H}iggs boson couplings to photons},
  {\emph{Sov. J. Nucl. Phys.} {\bfseries 30} (1979) 711}.

\bibitem{Marciano:2011gm}
W.J.~Marciano, C.~Zhang and S.~Willenbrock, \emph{Higgs decay to two photons},
  \href{https://doi.org/10.1103/PhysRevD.85.013002}{\emph{Phys. Rev. D}
  {\bfseries 85} (2012) 013002}
  [\href{https://arxiv.org/abs/1109.5304}{{\ttfamily 1109.5304}}].

\end{thebibliography}\endgroup

\end{document}